# Generalized Dirac oscillator in cosmic string space-time

Lin-Fang Deng[1], Chao-Yun Long[1,2]†, Zheng-Wen Long[1] and Ting Xu[1]

1) Department of Physics, Guizhou University, Guiyang 550025, China

2) Laboratory for Photoelectric Technology and Application, Guizhou University, Guiyang 550025, China

Abstract:

In this work, the generalized Dirac oscillator in cosmic string space-time is studied by replacing the momentum $p_\mu$ with its alternative $p_\mu + m\omega\beta f_\mu(x_\mu)$. In particular, the quantum dynamics is considered for the function $f_\mu(x_\mu)$ to be taken as cornell potential, exponential-type potential and singular potential. For cornell potential and exponential-type potential, the corresponding radial equations can be mapped into the confluent hypergeometric equation and hypergeometric equation separately. The corresponding eigenfunctions can be represented as confluent hypergeometric function and hypergeometric function. The equations satisfed by the exact energy spectrum have been found. For singular potential, the wave function and energy eigenvalue are given exactly by power series method.

## 1.Introduction

In quantum mechanics, there has been an increasing interest in finding the analytical solutions that play an important role for getting complete information about quantum mechanical systems [1-3]. The Dirac oscillator proposed in Ref. [4] is one of the important issues in this relativistic quantum mechanics recently. In this quantum model, the coupling proposed is introduced in such a way that the Dirac equation remains linear in both spatial coordinates and momenta, and recover the Schrödinger equation for a harmonic oscillator in the nonrelativistic limit of the Dirac equation [4-11]. As a solvable model of relativistic quantum mechanical system, the Dirac oscillator has many applications and has been studied extensively in different fieldssuch as high energy physics[12-15],condensed matter physics[16-18], quantum Optics[19-25] and mathematical physics[26-33] etc. On the other hand, the analysis of gravitational interactions with a quantum mechanical system has recently attracted a great deal attention and has been an active field of research[5-6,34-43].The study of quantum mechanical problems in curved space-time can be considered as a new kind of interaction between quantum matter and gravitation in the microparticle world. In recent years, the Dirac oscillator embedded in a cosmic string background has inspired a great deal of research such as the dynamics of Diracoscillator in the space-time of cosmic string[44-47], Aharonov-Casher effect on the Dirac oscillator[5,48], noninertial effects onthe Dirac oscillator in the cosmic string space-time[49-51]etc. It is worth mentioning that based on the coupling corresponding to the Dirac oscillator a new coupling into Dirac equation first has been proposed by Bakke *et al*[52]. and used in different fields[53-57]. This model is called the generalized Dirac oscillator which, in special case is reduced to ordinary Dirac oscillator. Inspired by the above work, the main aim of this paper is to analyze the generalized Dirac oscillator model with the interaction functions $f_\mu(x_\mu)$ taken as cornell potential, singular potential and exponential-type potential in cosmic string space-time and to find the corresponding energy spectrum and wave functions. This work is organized as follows. In sect.2, the new coupling is introduced in such a way that the Dirac equation remains linear in momenta, but not in spatial coordinates in a curved space-time. In sect.3, Sect.4 and Sect.5, we concentrate our efforts inanalytically solving the quantum systems with different function $f_\mu(x_\mu)$ and find the corresponding energy spectrum and spinors respectively. In sect.6, we make a short conclusion.

†e-mail: longchaoyun66@163.com

## 2、Generalized Dirac oscillator with a topological defect

In cosmic string space–time, the general form of the cosmic string metric in cylindrical coordinates read [41,42,44,58,59]

$$ds^2 = -dt^2 + d\rho^2 + \alpha^2\rho^2 d\varphi^2 + dz^2, \tag{1}$$

with $-\infty < (t, z) < +\infty$, $0 < \rho < +\infty$ and $0 < \varphi < 2\pi$. The parameter α is related to the linear mass density of string $\eta$ by $\alpha = 1 - 4\eta$ and runs in the interval $(0, 1]$. In the limit as $\alpha \to 1$ we get the line element of cylindrical coordinates. The Dirac equation in the curved space-time $(\hbar = c = 1)$ reads

$$[i\gamma^\mu(x)\partial_\mu - i\gamma_\mu(x)\Gamma_\mu(x) - m]\psi(t, x) = 0, \tag{2}$$

where the $\gamma^\mu$ matrices are the generalized Dirac matrices defining the covariant Clifford algebra $\{\gamma^\mu, \gamma^\nu\} = 2g^{\mu\nu}$, m is mass of the particle, and $\Gamma_\mu$ is the spinor affine connection. We choose the basis tetrad $e_a^\mu$ as

$$e_a^\mu = \begin{pmatrix} 1 & 0 & 0 & 0 \\ 0 & \cos\varphi & \sin\varphi & 0 \\ 0 & \dfrac{\sin\varphi}{\alpha\rho} & \dfrac{\cos\varphi}{\alpha\rho} & 0 \\ 0 & 0 & 0 & 1 \end{pmatrix}, \tag{3}$$

then in this representation the matrices $\gamma^\mu$[44] can be found to be

$$\gamma^0 = \gamma^t, \gamma^1 = \gamma^\rho = \gamma^1 \cos\varphi + \gamma^2 \sin\varphi,$$
$$\gamma^2 = \gamma^\varphi = -\gamma^1 \sin\varphi + \gamma^2 \cos\varphi, \tag{4}$$
$$\gamma^3 = \gamma^z, \quad \gamma^\mu \Gamma_\mu(x) = \frac{1-\alpha}{2\alpha\rho}\gamma^\rho.$$

It is well known that in both Minkowski spacetime and curved spacetime, usual Dirac oscillator can be obtained by the carrying out non-minimal substitution $p_\mu \to p_\mu + m\omega\beta x_\mu$ in Dirac equation where m and ω are the mass and oscillator frequency. In the following, we will construct the generalized oscillator in curved spacetime. To do this end, we can replace momenta $p_\mu$ in the Dirac equation of curved spacetime by

$$p_\mu \to p_\mu + m\omega\beta f_\mu(x_\mu), \tag{5}$$

where $f_\mu(x_\mu)$ are undetermined functions of $x_\mu$. It is to say that we introduce a new coupling in such a way that the Dirac equation remains linear in momenta but not in coordinates. In particular, in this work, we only consider the radial component the

non-minimal substitution

$$f_\mu(x_\mu) = (0, f_\rho(\rho), 0, 0). \tag{6}$$

By introducing this new coupling (6) into the equation(2) and with the help of the equation(4), in cosmic string space-time the eigenvalue equation of generalized Dirac oscillator can be written as

$$\left\{-i\gamma^t \partial_t + i\gamma^\rho \left(\partial_\rho - \frac{1-\alpha}{2\alpha\rho} + m\omega\rho f(\rho)\right) + \frac{i\gamma^\varphi \partial_\varphi}{\alpha\rho} + i\gamma^z \partial_z - m\right\}\psi = 0. \tag{7}$$

We choose the following ansatz

$$\psi = e^{-iEt + i(l+1/2-\Sigma^3/2)\varphi + ikz}\begin{pmatrix}\chi_1(\rho)\\ \chi_2(\rho)\end{pmatrix}, \tag{8}$$

then we have

$$\left[\alpha_1\left(\frac{d}{d\rho} + \frac{1}{2\rho} - m\omega\rho f(\rho)\right) - \frac{\lambda}{\rho}\alpha_2 - k\alpha_3\right]\left[\alpha_1\left(\frac{d}{d\rho} + \frac{1}{2\rho} + m\omega f(\rho)\right) - \frac{\lambda}{\rho}\alpha_2 - k\alpha_3\right]$$
$$\chi_1 = (E^2 - m^2)\chi_1, \tag{9}$$

$$\left[\alpha_1\left(\frac{d}{d\rho} + \frac{1}{2\rho} + m\omega f(\rho)\right) - \frac{\lambda}{\rho}\alpha_2 - k\alpha_3\right]\left[\alpha_1\left(\frac{d}{d\rho} + \frac{1}{2\rho} - m\omega f(\rho)\right) - \frac{\lambda}{\rho}\alpha_2 - k\alpha_3\right]$$
$$\chi_2 = (E^2 - m^2)\chi_2, \tag{10}$$

where

$$\alpha_1 = i(\sigma_1 \cos\varphi + \sigma_2 \sin\varphi),$$
$$\alpha_2 = -\sigma_1 \sin\varphi + \sigma_2 \cos\varphi, \quad \alpha_3 = \sigma_3. \tag{11}$$

It is straightforward to prove the following relations satisfied above matrices $\alpha_i$

$$\alpha_1^2 = -\alpha_2^2 = \alpha_3^2 = -1,$$
$$\alpha_1\alpha_2 = -\alpha_2\alpha_1 = i\sigma_1\sigma_2,$$
$$\alpha_1\alpha_3 = -\alpha_3\alpha_1 = i(\sigma_1\sigma_3\cos\varphi + \sigma_2\sigma_3\sin\varphi), \tag{12}$$
$$\alpha_3 = -\alpha_3\alpha_2 = -\sigma_1\sigma_3\sin\varphi + \sigma_2\sigma_3\cos\varphi.$$

With help of the equation (12) and simple algebraic calculus the equation (9)becomes

$$\left\{\partial_\rho^2 + \frac{1}{\rho}\partial_\rho - \frac{1}{\rho^2}\left[\frac{1}{4} + i\lambda\sigma_1\sigma_2 + \lambda^2\right]\right\}\chi_1$$
$$+ \left\{-2m\omega\frac{f(\rho)}{\rho}[ik\rho(\sigma_1\sigma_3\cos\varphi + \sigma_2\sigma_3\sin\varphi) + i\sigma_1\sigma_2\lambda]\right\}\chi_1$$

$$+\left\{m^2 + k^2 - E^2 + m\omega\frac{f(\rho)}{\rho} - m^2\omega^2 f^2(\rho)\right\}\chi_1 = 0. \qquad (13)$$

It is easy to prove the following relation [44]

$$i\sigma_1\sigma_2\lambda + ik\rho(\sigma_1\sigma_3\cos\varphi + \sigma_2\sigma_3\sin\varphi) = -2\vec{s}.\vec{L}, \qquad (14)$$

where $\vec{s} = \frac{\vec{\sigma}}{2}$. The eigenvalue of $\vec{s}.\vec{L}$ can be assumed as $(l + 1/2)/2\alpha$ and the equation (13) reads

$$\frac{d^2\chi_1}{d\rho^2} + \frac{1}{\rho}\frac{d}{d\rho}\chi_1 - \left[\frac{\lambda^2}{\rho^2} + \mu\frac{f(\rho)}{\rho} - m\omega\frac{df(\rho)}{d\rho} + m^2\omega^2 f^2(\rho)\right]\chi_1 + \nu\chi_1 = 0, \quad (*a)$$

where

$$\lambda = \left(\frac{l+1/2}{\alpha} - \frac{1}{2}\right), \quad \mu = \frac{-2(l+1/2)m\omega}{\alpha}, \quad \nu = E^2 - m^2 - k^2 ., \qquad (15)$$

For the component $\chi_2$, from equation (10) an analog equation can be also obtained

$$\frac{d^2\chi_2}{d\rho^2} + \frac{1}{\rho}\frac{d}{d\rho}\chi_2 - \left[\frac{\lambda^2}{\rho^2} + \mu\frac{f(\rho)}{\rho} + m\omega\frac{df(\rho)}{d\rho} + m^2\omega^2 f^2(\rho)\right]\chi_2 + \nu\chi_2 = 0, \quad (*b)$$

where

$$\lambda = \left(\frac{l+1/2}{\alpha} + \frac{1}{2}\right), \quad \mu = \frac{-2(l+1/2)m\omega}{\alpha}, \quad \nu = E^2 - m^2 - k^2. \qquad (16)$$

In particular, the equation (14) will reduced to the result obtained in Ref.[44] when the function f(ρ) is taken as f(ρ) = ρ. As we can see, the equation (*a) and thequation (*b) have the same form. Sowithout loss of generality in remaining parts of this work, our main tasks is onlyto solve the equation (*a) with different functions f(ρ) and find corresponding eigenvalue and eigenfunction. While with regard to the quation (*b), it is straightforward to obtain the corresponding solution.

3、 **The solution with f(ρ) to be cornell potential**

The cornell potential that consists of Coulomb potential and linear potential, has gotten a great deal of attention in particle physics and was used with considerable success in models describing systems of bound heavy quarks[60-62]. In cornell potential, the short-distance Coulombic interaction arises from the one-gluon exchange between the quark and its antiquark, and the long-distance interaction is included to take into account confining phenomena.

Now we let the function f(ρ) to be cornell potential

$$f(\rho) = a\rho - \frac{b}{\rho}, \tag{17}$$

where a and b are two constants. Substituting (17) into (*a) and (*b) leads to following equation

$$\frac{d^2\chi}{d\rho^2} + \frac{1}{\rho}\frac{d}{d\rho}\chi + \left[\frac{-\tau_1^2}{\rho^2} - \tau_2\rho^2 + \tau_3\right]\chi = 0, \tag{18}$$

where

$$\tau_1^2 = \lambda^2 - \mu b + m^2\omega^2 b^2 - \omega m b,$$
$$\tau_2 = m^2\omega^2 a^2,, \tag{19}$$
$$\tau_3 = \upsilon + 2abm^2\omega^2 - a\mu + m\omega a.$$

We make a change in variables $\xi = m\omega a\rho^2$ and then the equation(18) can be rewritten as

$$\xi\frac{d^2\chi}{d\xi^2} + \frac{d}{d\xi}\chi + \left[\frac{-\tau_1^2}{4\xi} - \frac{1}{4}\xi + \frac{\tau_3}{4ma\omega}\right]\chi = 0. \tag{20}$$

Taking account of the boundary conditions satisfied by the wave function χ, i.e., $\chi \propto \xi^{\tau_1/2}$ for $\xi \to 0$ and $\chi \propto e^{-\frac{\xi}{2}}$ for $\xi \to \infty$, physical solutions χ can be expressed as [44,60,63]

$$\chi = \xi^{|\tau_1|/2} e^{-\frac{\xi}{2}} g(\xi). \tag{21}$$

If we insert this wave function χ into Eq.(20), we have the second-order homogeneous linear differential equation in the following form:

$$\xi\frac{d^2 g}{d\xi^2} + (|\tau_1| + 1 - \xi)\frac{d}{d\xi}g + \left[\frac{\tau_3}{4ma\omega} - \frac{|\tau_1|}{2} - \frac{1}{2}\right]g = 0. \tag{22}$$

It is well known that the equation(22) is the confluent hypergeometric equation and it is immediate to obtain the corresponding eigenvalues and eigenfunctions

$$g(\xi) = F\left[-\left(\frac{\tau_3}{4ma\omega} - \frac{|\tau_1|+1}{2}\right), |\tau_1| + 1, \xi\right], \tag{23}$$

$$E_n^2 = \delta_1 - \frac{2(l+1/2)m\omega a}{\alpha} + 4ma\omega\left[n + \frac{1}{2} + \frac{|\delta_2|}{2}\right], \tag{24}$$

with

$$\delta_1 = m^2 + k^2 - 2abm^2\omega^2 + m\omega a,$$
$$\delta_2 = \sqrt{\lambda^2 - \mu b + b^2 m^2\omega^2 - m\omega b}., \tag{25}$$

In particular, if we assume that $\alpha = 1$, from equation(24), the energy levels of generalized Dirac oscillator with $f(\rho)$ to be cornell potential, in the absence of a topological defectcan obtained. In addition if we let $a = 1, b = 0$ in equation(24) the energy levels given here will bereduced to that one obtained in reference[44].

## 4. The solution with $f(\rho)$ to be singular potential

The investigation of singular potentials in quantum mechanics is almost as old asquantum mechanics itself and covers a wide range of physical and mathematical interest because the real world interactions were likely to be highly singular[64]. Thesingular potentials of $v(r) \propto \frac{1}{r^n}$ type, with $n \geq 2$ is of great current physical interest and is relevant to many problems such as the three-body problem in nuclear physics [65-66], point-dipole interactions in molecular physics[67], the tensor force between nucleons[68], and the interaction between a chargesand an induced dipole[69] respectively. Recently, in cosmic string background, singular inverse square potential with a minimal length had been studied[70].

Next let us take$f(\rho)$ to be singular inverse-square -type potential [71]

$$f_\rho(\rho) = a + \frac{b}{\rho} + \frac{c}{\rho^2}. \tag{26}$$

Substituting Eq. (26) into Eq. (*a), the corresponding radialequation reads

$$\left\{\frac{d^2}{d\rho^2} + \frac{1}{\rho}\frac{d}{d\rho} - \frac{\delta_1}{\rho} - \frac{\lambda^2 + \delta_2}{\rho^2} - \frac{\delta_3}{\rho^3} - \frac{\delta_4}{\rho^4} + \gamma\right\}\chi = 0, \tag{27}$$

where
$$\gamma = \upsilon - m^2 a^2 \omega^2$$
$$\delta_1 = 2abm^2\omega^2 + a\mu,$$
$$\delta_2 = \mu b + m\omega b + b^2 m^2 \omega^2 + 2acm^2\omega^2,, \tag{28}$$
$$\delta_3 = \mu c + 2m\omega c + 2bcm^2\omega^2,$$
$$\delta_4 = c^2 m^2 \omega^2.$$

It is obvious that (27) has the same mathematical structure with the Schrödinger equation of fourth-order inverse-potential in Ref.[72]. So the (27) can solved by power series method.

We look for an exact solution of (27) via the following ansatz to the radial wave function[72,73,74]

$$\chi = \Theta(\rho)\exp[g(\rho)], \tag{29}$$

$$g(\rho) = -\frac{\delta_1}{2\rho} - \frac{\delta_2}{2}\rho - \frac{\delta_3}{2}\log\rho.$$

Thence, Eq.(27) can be transformed into the following form

$$\left\{\frac{d^2}{d\rho^2} + \left[-\delta_1 + \frac{1-\delta_3}{\rho} + \frac{\delta_2}{\rho^2}\right]\frac{d}{d\rho} + \gamma + \frac{\delta_1^2}{4} + \frac{\delta_1(\delta_3-3)}{2\rho} + \frac{\delta_3^2-4\delta_2-2\delta_1\delta_2-4\lambda^2}{4\rho^2} + \frac{-\delta_2(1+\delta_3)-\delta_3}{2\rho^3} + \frac{\delta_2^2-4\delta_4}{4\rho^4}\right\}\Theta(\rho) = 0 .$$ 
(30)

Now we take $\Theta(\rho)$ in following series form

$$\Theta(\rho) = \sum_{n=0}^{\infty} a_n \rho^{n+\lambda+\frac{1}{2}}, a_0 \neq 0 , \quad a_1 \neq 0 .$$ 
(31)

Substituting equation(31) into equation (30) gives rise following equation

$$\sum_{n=0}^{\infty} a_n \left\{-\left[\delta_1\left(n+\lambda+2-\frac{\delta_3}{2}\right)\right]\rho^{n+\lambda-\frac{1}{2}} + \left[\gamma + \frac{\delta_1^2}{4}\right]\rho^{n+\lambda+\frac{1}{2}} \right.$$

$$+ \left[\frac{(2n+2\lambda+1)(2n+2\lambda+1-2\delta_3)}{4} - \frac{\delta_2(\delta_1+2)}{2} - \lambda^2 + \frac{\delta_3^2}{4}\right]\rho^{n+\lambda-\frac{3}{2}}$$

$$\left. - \left[\delta_1\left(n+\lambda+2-\frac{\delta_3}{2}\right)\right]\rho^{n+\lambda-\frac{1}{2}} + \left[\gamma + \frac{\delta_1^2}{4}\right]\rho^{n+\lambda+\frac{1}{2}}\right\} = 0 .$$ 
(32)

To make equation(32) be valid for all values of $\rho$, the coefficients of each term of the polynomial of $\rho$ must be equal to zero separately. We, therefore, obtain

$$2(\delta_1 + 2)(n + \lambda + 2) = 4\lambda^2 - 9 - 4,$$ (33a)

$$\delta_2 = -(n + \lambda + 2),$$ (33b)

$$\delta_3 = 2(n + \lambda + 2) ,$$ (33c)

$$2\delta_4 = (n + \lambda + 2)^2 ,$$ (33d)

$$\gamma = -\frac{\delta_1^2}{4} .$$ (33e)

Using equations (*a),(33a) and (33e) and after simple algebraic calculation, the corresponding energy can be written as

$$E_n^2 = m^2 + k^2 + m^2 a^2 \omega^2 - \frac{1}{16}\left(\frac{4\lambda^2-9}{n+\lambda+2}\right)^2.$$ (34)

The general radial wave functions corresponding to the energy spectra given in equation (34) are

$$\Theta(\rho) = \sum_{n=0}^{\infty} a_n \rho^{n+\lambda+\frac{1}{2}} \exp\left[-\frac{\delta_1}{2\rho} - \frac{\delta_2}{2}\rho - \frac{\delta_3}{2}\log\rho\right].$$ (35)

With the help of equations (32) and (33a)-(33e), the expansion coefficients $a_n$ in

equation(35) satisfy the following recursion relation [72]

$$a_{n+1} = \frac{4\lambda^2 - 9 - 4(n+\lambda+2)}{2(n+\lambda+2)^2} a_{n-1} .\qquad(36)$$

From the recursion relation(36) we can determinethecoefficients$a_n$ $(n \neq 0,1)$ of the power series in terms of $a_0$ and $a_1$. In addition the above recursion relation implies that equation (35) yields one solution as a power series in even powers of ρ and another in odd powers of ρ.

In addition, the equation (27) can be also mapped to the double-confluent Heun equation by appropriate function transformation[75]. So when f(ρ) is taken as singular inverse-square-type potential, the solutions of the equation(27) can been also given by the solution of the double-confluent Heun equation[75,76].

## 5、The solution with f(ρ) to be exponential-type potential

The exponential-type potentials are very important in the study of various physical systems, particularly for modeling diatomic molecules. The typical exponential-type potentials include Eckart potentials[77], the Morse potential[78,79], the Wood–Saxon potential[80] and Hulthén potential[81,82] etc. The research work on the Dirac equation with the above potential is mainly concentrated on Minkowski time and space. However, it has been noticed recently that it is also interesting to stusy this kind quanstum systems in a cosmic string background [83]. In this section we will take the f(ρ) as exponential-type functionand solve the corresponding Dirac equation in cosmic string space-time.

As is known to all，the Dirac equation and Schrödinger equation have been studied by resorting different methods. A usual way is transforming the eigenvalue equation of quantum system considered into a solvable equation via suitable variable substitutions and function transformations[84-86]. In order to obtain solution forf(ρ) being exponential-type potential, we firstly consider the following linear second-order differential equation

$$x^2(1-x)^2 \frac{d^2y}{dx^2} + x(1-x)^2 \frac{dy}{dx} - (\mathcal{L}_1 + \mathcal{L}_2 x - \mathcal{L}_3 x^2)y = 0,\qquad(37)$$

where $\mathcal{L}_i, (i = 1,2,3)$ are constants. It is known that singular points of a differential equation determine the form of solutions. In this equation, there are two singular points, i.e., $x = 0$ and $x = 1$. In order to remove these singularities and get physically acceptable solutions we use the following ansatz:

$$y = x^{\Omega}(1-x)^{\Lambda} R(x). \tag{38}$$

where $\Lambda$ and $\Omega$ are two real parameters. Further we make this two parameters to satisfy following relationships

$$\Omega = \pm\sqrt{\mathcal{L}_1}, \Lambda = \frac{1}{2}\left[1 \pm \sqrt{1 - 4(\mathcal{L}_3 - \mathcal{L}_2 - \mathcal{L}_1)}\right], \tag{39}$$

and by substitution the equation (38) to the equation (37), the differential equation for $\chi(x)$ can be written as

$$(1-x)x\frac{\partial^2}{\partial x^2}R(x) + [2\Omega + 1 - (2\Omega + 2\Lambda + 1)]x\frac{\partial}{\partial x}R(x)$$
$$-(\Omega + \Lambda + \Delta)(\Omega + \Lambda - \Delta)R(x) = 0, \tag{40}$$

with $\Delta = \pm\sqrt{-\mathcal{L}_3}$. In other words, the equation(37) is reduced to the well-known hypergeometric equation, when the condition(39) be imposed. Making use of the boundary conditions at $r = 0$ and $r = \infty$[86, 87], we can find the equation obeyed by the energy eigenvalue :

$$\Omega + \Lambda + \Delta = -n, \tag{41}$$

and the corresponding eigenfunctions is given in terms of the Gauss hypergeometric functions below

$$R(x) = AF(\tau_1, \tau_2; 1 + 2\Omega; x), \tag{42}$$
$$\tau_1 = \Omega + \Lambda + \Delta, \qquad \tau_2 = \Omega + \Lambda - \Delta,$$

where A is normalization constant. Next , we will use the results given here to obtain the solutions of Dirac equation exponential-type interaction in cosmic string space-time. As a direct application of the above method, let us take the function $f(\rho)$ to be as Yukawa potential, Hulthén-Type potential and generalized Morse potential respectively.

**Case 1. $f(\rho)$ being Yukawa potential**

In Yukawa meson theory, the Yukawa potential firstly was introduced to describe the interactions between nucleons[88]. Afterwards, it has been applied to many different areas of physics such as high-energy physics[89,90], molecular physics[91] and plasma physics[92]. In recent years, the considerable efforts have also been made to study the bound state solutions by using different methods.

Now let us choose $f(\rho)$ to be Yukawa potential

$$f(\rho) = \frac{a}{\rho} e^{-\beta\rho}, \tag{43}$$

then the equation (14a) takes the form

$$\frac{d^2\chi}{d\rho^2} + \frac{1}{\rho}\frac{d\chi}{d\rho} - \left[\frac{\lambda^2}{\rho^2} + \frac{m\omega a}{\rho^2} e^{-\beta\rho} + \frac{\mu a + am\omega\beta}{\rho} e^{-\beta\rho} + \frac{a^2 m^2 \omega^2}{\rho^2} e^{-2\beta\rho}\right]\chi + \nu\chi = 0. \tag{44}$$

However, the radial equation(44) cannot accept exact solution due to the presence of the centrifugal term[85]. In order to find analytical solution, we have to usesome approximation approaches for the centrifugal term potential. Following reference [86], the approximation for the centrifugal term reads:

$$\frac{1}{\rho^2} \approx \frac{\beta^2}{(1-e^{-\beta\rho})^2}, \frac{1}{\rho} \approx \frac{\beta}{1-e^{-\beta\rho}}. \tag{45}$$

It's worth mentioning that the above approximations are valid for $\beta\rho \ll 1$[86]. So If we make the controlparameter $\beta$ small enough, then we can guarantee that the above approximations in equation (45)holds for larger values $\rho$. In other words, this approximation(45) is valid in our work.

Using the approximation in Eq. (45),and setting

$$\chi = \frac{1}{\sqrt{\rho}} \Theta(\rho), \quad x = e^{-\beta\rho}, \tag{46}$$

in the new function $\chi$ and new variable$x$,the equation (44) becomes

$$x^2(1-x)^2 \frac{d^2\Theta}{dx^2} + x(1-x)^2 \frac{d\Theta}{dx} - (\mathcal{L}_1 + \mathcal{L}_2 x - \mathcal{L}_3 x^2)\Theta = 0, \tag{47}$$

where

$$\begin{aligned}
\mathcal{L}_1 &= -\lambda^2\beta^2 + m^2 + k^2 - E^2, \\
\mathcal{L}_2 &= -m\omega a\beta^2 - a\mu\beta + 2m^2 + 2k^2 - 2E^2, \\
\mathcal{L}_3 &= m^2\omega^2 a^2 - a\mu\beta - am\omega\beta^2 - m^2 - k^2 + E^2.
\end{aligned} \tag{48}$$

Comparing equation(47)with(37) and using the results given in the equation (41) and (42),it is not difficult to find the equation obeyed byeigenvalues and eigenfunctions and they can be given respectively

$$E_n^2 - q_1 - \sqrt{q_2 - E_n^2} + \left[n + \frac{1}{2}\left(1 + \sqrt{1 + q_3 - 16E_n^2}\right)\right] = 0, \tag{49}$$

$$\Theta(\rho) = Ae^{-\beta\Omega\rho}(1 - e^{-\beta\rho})^{\Lambda} F(\tau_1, \tau_2; 1 + 2\Omega; e^{-\beta\rho}),$$

Where

$$q_1 = \lambda^2\beta^2 - m^2 - k^2 , q_2 = m^2 + k^2 + a\mu\beta + am\omega\beta^2 - m^2\omega^2 a^2 ,$$

$$\sigma_3 = 4(4m^2 + 4k^2 - am^2\omega^2) , \tau_1 = \Omega + \Lambda + \Delta, \tau_2 = \Omega + \Lambda - \Delta. \quad (50)$$

**Case 2. f($\rho$) being Hulthén-Type potential**

In this section, we we are interested in considering the Hulthén potential that describe the interaction between two atoms and has been used in different areas of physics and attracted a great of interest for some decades[81,82,93]. Next we take the interaction function f($\rho$) being Hulthén-Type potential

$$f(\rho) = a + \frac{be^{-\beta\rho}}{1-e^{-\beta\rho}}, \quad (51)$$

where a, b and $\beta$ are real constants. Inserting the equations (45) and (51) into Eq. (*a), then the equation (*a) can written as

$$\frac{d^2\chi}{d\rho^2} + \frac{1}{\rho}\frac{d\chi}{d\rho} - \left[\frac{\lambda^2}{\rho^2} + \frac{\mu a}{\rho} + a^2m^2\omega^2 + \left(\frac{\mu b}{\rho} + 2abm^2\omega^2\right)\frac{e^{-\beta\rho}}{1-e^{-\beta\rho}}\right]\chi$$

$$+ (b^2m^2\omega^2 e^{-\beta\rho} + m\omega b\beta)\frac{e^{-\beta\rho}}{(1-e^{-\beta\rho})^2}\chi + \nu\chi = 0. \quad (52)$$

In the same way as in previous section, Taking into consideration approximation(45) for the centrifugal term, and using the variable transformation $x = e^{-\beta\rho}$ and function transformation $\chi = \frac{1}{\sqrt{\rho}}\Theta(\rho)$ the equation(51)changes

$$x^2(1-x)^2\frac{d^2\Theta}{dx^2} + x(1-x)^2\frac{d\Theta}{dx} - (\mathcal{L}_1 + \mathcal{L}_2 x - \mathcal{L}_3 x^2)\Theta = 0, \quad (53)$$

where

$$\mathcal{L}_1 = \lambda^2\beta^2 + m^2\omega^2 a^2 + \beta\mu a + m^2 + k^2 - E^2,$$

$$\mathcal{L}_2 = m\omega b\beta + (b-a)(\mu\beta - 2m\omega) + 2m^2 + 2k^2 - 2E^2, \quad (54)$$

$$\mathcal{L}_3 = -m^2\omega^2(a-b)^2 - m^2 - k^2 + E^2.$$

With the help of equations (38),(41) and (42), the solutions for f($\rho$) being Hulthén-Type potential can beeasily obtained and the corresponding eigenvalues and eigenfunctions respectively

$$E_n^2 - q_4 - \sqrt{q_5 - E_n^2} + \left[n + \frac{1}{2}\left(1 + \sqrt{1 + q_6 - 16E_n^2}\right)\right] = 0, \quad (55)$$

$$\Theta(\rho) = Ae^{-\beta\Omega\rho}(1 - e^{-\beta\rho})^{\Lambda} F(\tau_1, \tau_2; 1 + 2\Omega; e^{-\beta\rho}), \tag{56}$$

where

$$q_4 = \lambda^2\beta^2 + m^2\omega^2 a^2 + \beta\mu a + m^2 + k^2,$$
$$q_5 = m^2\omega^2(a-b)^2 + m^2 + k^2,$$
$$q_6 = 4(m^2 + k^2) + 4m^2\omega^2(2a^2 - 2ab + b^2) + 4\beta^2\lambda^2 + 4\beta\mu a$$
$$\quad + 4m\omega b\beta + 4(a-b)(\mu\beta - 2m\omega), \tag{57}$$
$$\tau_1 = \Omega + \Lambda + \Delta, \quad \tau_2 = \Omega + \Lambda - \Delta.$$

### Case 3. $f(\rho)$ being generalized Morse potential

The morse potential[78-79] as an important molecular potential describes the interaction between two atoms. We choose the interaction function $f(\rho)$ being generalized morse potential

$$f(\rho) = \frac{a}{\rho^2}\left(e^{-\beta\rho} - e^{-2\beta\rho}\right). \tag{58}$$

As before, substitution of the form （59）into Eqs.（*a）and straightforward calculation leads to the equation

$$\frac{d^2\chi}{d\rho^2} + \frac{1}{\rho}\frac{d\chi}{d\rho} - \left[\frac{\lambda^2}{\rho^2} + \frac{\mu a + 2m\omega a\mu}{\rho^3}\left(e^{-\beta\rho} - e^{-2\beta\rho}\right) - \frac{a\beta m\omega}{\rho^2}\left(2e^{-2\beta\rho} - e^{-\beta\rho}\right)\right]\chi$$
$$+ \frac{a^2 m^2 \omega^2}{\rho^4}\left(e^{-\beta\rho} - e^{-2\beta\rho}\right)^2 \chi + \nu\chi = 0. \tag{59}$$

Letting $x = e^{-\beta\rho}$ and $\chi = \frac{1}{\sqrt{\rho}}\Theta(\rho)$, the above differential equation (59) changes into the form

$$x^2(1-x)^2 \frac{d^2\Theta}{dx^2} + x(1-x)^2 \frac{d\Theta}{dx} - (\mathcal{L}_1 + \mathcal{L}_2 x - \mathcal{L}_3 x^2)\Theta = 0, \tag{60}$$

Where

$$\mathcal{L}_1 = \lambda^2\beta^2 + m^2 + k^2 - E^2,$$
$$\mathcal{L}_2 = \mu a(1 + 2m\omega) + 2am\omega\beta - 2m^2 - 2k^2 + 2E^2, \tag{61}$$
$$\mathcal{L}_3 = E^2 - m^2\omega^2 a^2 + 2m\omega a\beta - m^2 - k^2.$$

It is easy to see that the differential equation(60) is also similar to the equation (37). So again according to the quantization condition (40) the corresponding expression of eigenvalues can be written as

$$E_n^2 - q_7 - \sqrt{q_8 - E_n^2} + \left[n + \frac{1}{2}\left(1 + \sqrt{1 + q_9 - 16E_n^2}\right)\right] = 0, \tag{62}$$

$$q_7 = \lambda^2\beta^2 + m^2 + k^2,$$

$$q_8 = m^2\omega^2 a^2 - 2m\omega a\beta + m^2 + k^2,$$

$$q_9 = 4[\mu a(1 - 2m\omega) - 2m\omega a\beta + a^2 m^2 \omega^2 + \lambda^2 \beta^2].$$

The wave function in this case read

$$\Theta(\rho) = Ae^{-\beta\Omega\rho}(1 - e^{-\beta\rho})^\Lambda F(\tau_1, \tau_2; 1 + 2\Omega; e^{-\beta\rho}), \tag{63}$$

where

$$\tau_1 = \Omega + \Lambda + \Delta, \quad \tau_2 = \Omega + \Lambda - \Delta. \tag{64}$$

The above results show that the radial equation of the generalized Dirac oscillator with interaction function $f_\mu(x_\mu)$ to be taken as the exponential-type potential can be mapped into the well-known hypergeometric equation and the analytical solutions can have been found.

## 6. Conclusion

In this work, the generalized Dirac oscillator has been studied in the presence of the gravitational fields of a cosmic string. The corresponding radial equation of generalized Dirac oscillator is obtained. In our generalized Dirac oscillator model, we take the interaction function $f_\mu(x_\mu)$ to be as cornell potential, Yukawa potential, generalized morse potential, Hulthén-Type potential and singular potential respectively. By solving the corresponding wave equations the corresponding energy eigenvalues and the wave functions have been obtained and we have showed how the cosmic string, leads to modifications in the spectrum and wave function. Based on consideration that Dirac oscillator has been studied extensively in high energy physics, condensed matter physics, quantum Optics, mathematical physics and even in connection with Higgs symmetry it also makes sense to generalize the generalized Dirac oscillator to these fields.

**Acknowledgements**

This work is supported by the National Natural Science Foundation of China

(Grant Nos.,11565009).